\documentclass[12pt]{elsarticle}
\usepackage{graphics,graphicx}
\journal{Physica A}
\begin{document}

\title{Effect of localized loading on failure threshold of fiber bundles}



\author[label1]{Soumyajyoti Biswas} 
\ead{soumyajyoti.biswas@ds.mpg.de}
\author[label2]{Parongama Sen}
\ead{psphy@caluniv.ac.in}
%\email[]{soumyajyoti.biswas@ds.mpg.de}
%\author{Lucas Goehring}
%\email[]{lucas.goehring@ds.mpg.de}
%\thanks{}
\address[label1]{Max Planck Institute for Dynamics and Self-Organization, Am Fassberg 17, G\"{o}ttingen-37077, Germany.}
\address[label2]{Department of Physics, University of Calcutta, 92 Acharya Prafulla Chandra Road, Kolkata 700009, India.}
%\ead[label1]{soumyajyoti.biswas@ds.mpg.de}
%\ead[label2]{psphy@caluniv.ac.in}

\date{\today}

\begin{abstract}
\noindent  We investigate the global failure threshold of an interconnected set of elements, when a finite fraction of
the elements initially share an externally applied load. The study is done under the framework of random fiber bundle model, where
the fibers are linear elastic objects attached between two plates. The failure threshold of the system varies non-monotonically
with the fraction of the system on which the load is applied initially, provided the load sharing mechanism following a local
failure is sufficiently wide. In this case, there exists a finite value for the initial loading fraction, for which the damage on the system will
be maximum, or in other words the global failure threshold will be minimum for a finite value of the initial loading fraction. 
This particular value of initial loading fraction, however, goes to
zero when the load sharing is sufficiently local. Such crossover behavior, seen for both one and two dimensional versions of the model,
can give very useful information about stability of interconnected systems with random failure thresholds. 
\end{abstract}
%\pacs{89.75.Da, 64.60.av, 62.20.M-}
%\keywords{Tomlinson's Model, Static Friction, Cantor Set, Fractal Overlap}

\maketitle

\section{Introduction}    
An interconnected set of elements sharing a load is a common situation arising in diverse contexts such as the
disordered solids under stress \cite{book1,book2,book3,wood,cliff,land}, grids carrying current \cite{power1,power2},  network of computers sharing a task, 
network of roads carrying traffic \cite{traffic1,traffic2} etc. The catastrophic failure point in such systems, while mostly an undesirable 
situation, is a crucial factor in fixing the operating points and thereby limiting the resources and functionality of 
the relevant systems. A set of elements, or fibers, having random failure thresholds and fixed between two rigid plates, is a prototype model to study the breakdown
properties of a broad category of systems such as these, under simplifying yet informative assumptions. 

The so called fiber bundle model was introduced in the textile industry \cite{first} to model the strength distributions of cloths. Since then it has found
wide spread applications in systems with varying degree of complexities that still has the underlying basic dynamics of threshold
activated breakdown \cite{daniels}. While the individual fibers are often assumed to have a linear stress-strain relation with an irreversible breakdown
beyond a threshold, the overall response of the system is non-linear. Depending upon the distribution function of the individual failure
thresholds and the range of load redistribution following a local failure, the system can show nucleation driven extreme statistics to
random percolative failure through avalanche dynamics \cite{phoenix1,harlow,sornette,pre17}. 
The system size dependence of these response statistics, limiting cases of very strong or weak disorders in the system
and the range of load sharing etc are some of the important questions that are still being actively investigated \cite{pre17,biswas15,kun17}.

 In this work, however, we look
back at the prediction of failure threshold under a constant total load, the original question of the model. In most studies, the application of the initial load in fiber bundles
is uniform. Under that condition, the catastrophic failure threshold and its system size dependence are well studied and understood \cite{fbm_book,rmp1}. However,
much less attention was paid to the systems where the loading may not be uniform at the outset (see e.g. \cite{roy18}). This is, however, a very common
situation that can arise in all the examples mentioned above. For example, going back to the origin of the model, 
a constant load can either be applied uniformly on the lower plate supported by the fibers or can be distributed between a series of loading points.
In power grids the loading is known to be non-uniform (see e.g. \cite{power_load}), similar situation is true for traffics on the road, computer network with redundancies and so on. 
Such non-uniformity of the initial loading can
have very significant effect on the global failure threshold of the system. Here we show how the failure threshold, or the overall 
load carrying capacity of the system, varies with the fraction of system where the load is initially applied.   
Interestingly, we find the variation to be non-monotonic. This implies that in case of partial loading, certain fractions are to be 
avoided if the goal is to increase the overall failure thresholds. In other cases, where fracturing is desirable (e.g. hydraulic fracture in oil extraction),
such fractions are to be targeted to achieve maximum fracture.
  
In the following, we first investigate the fiber bundle model in the mean field limit, where the initial loads are applied to a 
finite fraction of the system. Even in this simple limit, we see that given a fixed total load, the damage on the system is maximum 
when a finite fraction (between $1/N$, $N$ being the system size, and 1) of it bears the initial load. In other words, the failure threshold
is minimum for that fraction. This point of minimum threshold is special and has different scaling of the fluctuation of the critical load than 
in other points both above and below this fraction. The mean field limit has some analytical tractability and hence can bring some insights to the dynamics. 
We then go over to the more realistic situations of having a finite compliance of the bottom plate \cite{stormo12}, or in other words a power law 
load sharing in the system \cite{hidalgo02, biswas16}, following a local breakdown. This is done for both the one dimensional and two dimensional versions of
the model. In both cases, for sufficiently wide load redistribution rule, we recover the non-monotonicity of the damage fraction 
(and of the critical threshold), which is retained up to certain degree of load redistribution range. For very localized load redistribution,
however, this non monotonicity vanishes and failure threshold becomes a monotonically increasing function of the initially loaded
fraction.

\section{Model}
In its original form, the fiber bundle model is viewed as a set of $N$ fibers fixed between two plates. The plates are either pulled apart (strain
controlled dynamics) or a fixed load is attached to the bottom plate (stress controlled dynamics). The individual fibers are linear elastic
and each of them can fail irreversibly once their failure thresholds are exceeded. The failure thresholds $\sigma_{th}^i$ are drawn randomly from
some probability distribution. Once a fiber fails, its share of load is then redistributed among the remaining intact fibers. The collective behavior 
of the system is surprisingly rich. It depends mainly on the properties of the distribution function the failure thresholds are chosen from and the way 
in which the load is shared between the remaining intact fibers. In this work, we will use a uniform threshold distribution between 
[0:1]. The load sharing mechanism will be varied. Specifically, we will study the mean field limit, where the load 
of a failed fiber is shared equally between all remaining intact fibers, and also the power law load sharing, where a fiber at site $r$ will have a load
share proportional to $1/|r-r^{\prime}|^{\alpha}$ when a fiber at site $r^{\prime}$ fails. Of course, the limit $\alpha\to 0$ is the mean field limit 
and $\alpha\to\infty$ is completely local load sharing limit. In practice, depending on the spatial dimension of the problem, which is either one or two here,
there is crossover value $\alpha^*$ for which the behavior of the model crosses over from mean field to local load sharing limit. 

The crucial difference in this work is that the load is initially applied to $p$ fraction of the fibers. The subsequent dynamics in the simulation is the same as is usually followed
in fiber bundle models i.e. the fibers having failure thresholds below the applied load on it are broken, the load carried by those fibers are then redistributed on the remaining surviving fibers depending on their distance from the broken fibers in case of power law load redistributions mentioned above or uniformly on all the fibers in case of mean field. The redistribution can trigger further breaking of the fibers and so on. For a given total load on the whole system, the system can either be stable in a state where all the surviving fibers carry a load below their respective thresholds or all the fibers are broken. The critical load for which the system just survives complete breakdown, is the critical point of the system. The simple modification of applying the load to a finite fraction of fibers initially, however, has a profound effect on the dynamics, stability and critical fluctuations of the system, which we shall investigate in the following for the
mean field and both one and two dimensional cases with power law load sharing.

%%%%%%%%%%%%%%%%%%%%%%%%%%%%%%%%%%%%%%%%%%%%%%%%%%%%%%%%%%%%%%%%%%%%%%%%%%%%%%%%%%%%%%%%%%%%%%%%%%%%%%%%%%%%%%%%%%%%%%%%%%%%%%%%%%%%%%%%%
\begin{figure}[tbh]
\centering 
%\captionsetup{justification=raggedright}
\includegraphics[height=9cm]{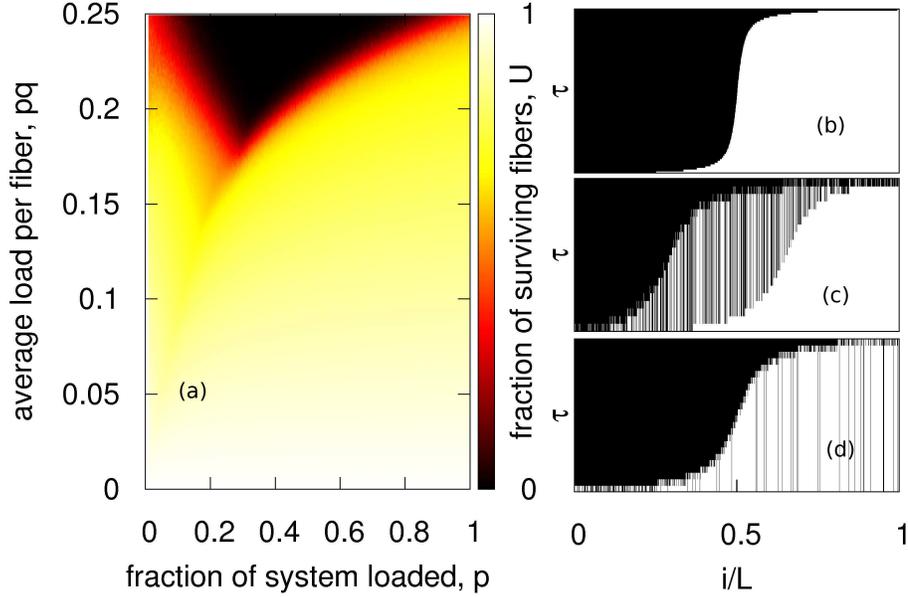}
   \caption{The surviving fraction of fibers are shown in (a) as a function of fraction of
system loaded ($p$) and total load on the system ($pq$). The critical line (between the light and dark shaded regions), beyond which no fiber
survives, is non-monotonic with $p$. The right hand sides figures (b-d) show the order of breaking of the 
fibers, when they are arranged in the ascending order of their failure thresholds and the system is
critically overloaded. The entire system is represented by surviving (white) broken (black) fibers as a function of the number of
load redistribution steps ($\tau$).  From top to bottom (b,c,d), the values of $p$ are 1.0, 0.6 and 0.2.}
\label{pd_fbm_partialload}
\end{figure}
%%%%%%%%%%%%%%%%%%%%%%%%%%%%%%%%%%%%%%%%%%%%%%%%%%%%%%%%%%%%%%%%%%%%%%%%%%%%%%%%%%%%%%%%%%%%%%%%%%%%%%%%%%%%%%%%%%%%%%%%%%%%%%%%%%%%%%%%%%
\section{Results} 
\subsection{Mean field model}
We begin with the simplest case of the mean field version of the model, where following the failure of one fiber, the load carried by that fiber
is uniformly redistributed among all the other remaining fibers. Now, if we assume $p$ fraction of the system is initially loaded with $q$ load per fiber each,
then the total load on the system is $pqN$. In Fig. \ref{pd_fbm_partialload}(a) we plot the fraction of surviving fibers, with $p$ and $pq$, 
i.e. going horizontally, the plot shows the fraction 
of fibers surviving for a given applied load when the initial loading fraction is varied. The dark-shaded region in the plot is where all the fibers have broken.
Separating that region is the critical line, which shows a dip for $p^*\approx 0.3$. The two limiting cases, $p\to 0$ and $p\to 1$ both show the usual critical
force $\sigma_c=1/4$. While the $p=1$ limit is well known, for $p\to 0 (1/N)$, the entire load is concentrated in one fiber, forcing it to break and the total force is now 
uniformly redistributed among the rest $N-1$ fibers. Hence the critical load here is only different from $p=1$ limit by an order $1/N$ correction. 

%%%%%%%%%%%%%%%%%%%%%%%%%%%%%%%%%%%%%%%%%%%%%%%%%%%%%%%%%%%%%%%%%%%%%%%%%%%%%%%%%%%%%%%%%%%%%%%%%%%%%%%%%%%%%%%%%%%%%%%%%%%%%%%%%%%%%%%%%
\begin{figure}[tbh]
\centering 
%\captionsetup{justification=raggedright}
\includegraphics[height=7cm]{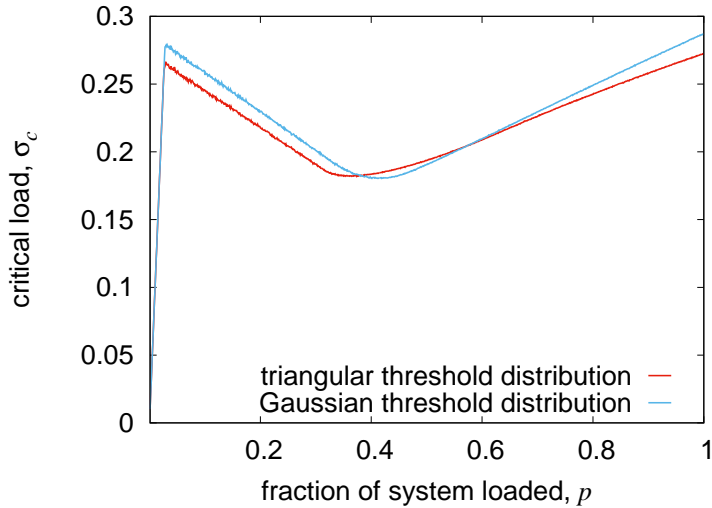}
   \caption{The critical load is plotted as a function of the fraction of initially loaded fibers $p$ when the threshold distributions of the 
fibers are Gaussian and triangular, with center at 0.5. The non-monotonic variation of the critical load is seen, like in the case of 
uniform threshold distribution mentioned above.}
\label{gaussian}
\end{figure}
%%%%%%%%%%%%%%%%%%%%%%%%%%%%%%%%%%%%%%%%%%%%%%%%%%%%%%%%%%%%%%%%%%%%%%%%%%%%%%%%%%%%%%%%%%%%%%%%%%%%%%%%%%%%%%%%%%%%%%%%%%%%%%%%%%%%%%%%%%
In fact, the part $p<p^*$ can be understood by extending the above logic. When $p$ is small, the initial load per fiber value is large, making sure every fiber
that had the initial load, breaks. In that case, the total load $pq$ is uniformly redistributed on $1-p$ remaining unloaded fraction. This is as if the usual uniform loading
 case, with a reduced system size $(1-p)N$. The phase boundary is, therefore, 
\begin{equation}
\frac{p_cq_c}{1-p_c}=\frac{1}{4},
\end{equation} 
giving, $4p_cq_c=1-p$, which the equation for the left part of the phase boundary. Now, this logic can be valid to the point where all initially loaded
fibers break i.e. $q_c\ge 1$, giving $p^*=1/5$. However, this is a rather strict condition for $p^*$, as the initially loaded system could also break in 
subsequent steps of redistribution. This is a lower bound for $p^*$. The estimate for $p^*$ from simulations is slightly higher. 
For this region ($0<p<p^*$), the order of breaking of the fibers is monotonic (except for the first step). The weaker fibers break first and the strongest
fibers survive till the end (see Fig. \ref{pd_fbm_partialload}(b-d)), which is also the case for usual fiber bundle model ($p=1$). 

%%%%%%%%%%%%%%%%%%%%%%%%%%%%%%%%%%%%%%%%%%%%%%%%%%%%%%%%%%%%%%%%%%%%%%%%%%%%%%%%%%%%%%%%%%%%%%%%%%%%%%%%%%%%%%%%%%%%%%%%%%%%%%%%%%%%%%%%%
\begin{figure}[tbh]
\centering 
%\captionsetup{justification=raggedright}
\includegraphics[height=7cm]{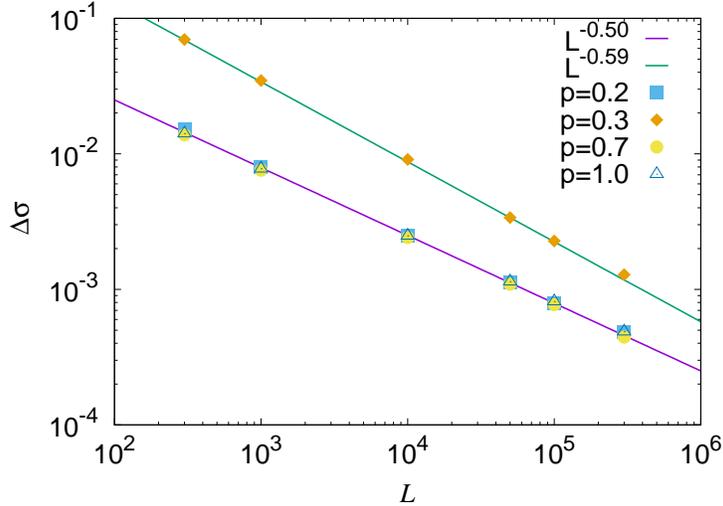}
   \caption{The rms fluctuations of the critical load varies with system sizes for different values of $p$. For $p=p^*\approx 0.3$ the scaling is
significantly different (with exponent $0.59\pm 0.02$) than that for other values of $p$ (exponent 0.5, exactly know in $p=1$ limit).}
\label{crpt_fluctuations}
\end{figure}
%%%%%%%%%%%%%%%%%%%%%%%%%%%%%%%%%%%%%%%%%%%%%%%%%%%%%%%%%%%%%%%%%%%%%%%%%%%%%%%%%%%%%%%%%%%%%%%%%%%%%%%%%%%%%%%%%%%%%%%%%%%%%%%%%%%%%%%%%%

The point $p=p^*\approx 0.3$ is special in the sense that the sequence of fibers breaking does not correspond to the sequence of their respective thresholds 
any more. This is because, some of the initially loaded strong fibers can sustain the first loading, but cannot support the subsequent redistributions. 
This effect is strongest at $p=p^*$ and can therefore have consequences in the critical fluctuation of the dynamics, particularly the failure thresholds.
We measured the fluctuation of the critical threshold $\Delta\sigma_c=\langle (\sigma^{(x)}-\langle \sigma_c\rangle)^2 \rangle^{1/2}_x$ for different $p$ values,
where the angular brackets denote the average over ensemble and $x$ is the ensemble index.
While for all other $p$ values $\Delta \sigma_c \sim L^{0.5}$, for $p=p*$, $\Delta\sigma_c \sim L^{-0.59\pm 0.02}$ for three orders of magnitude in the system sizes (see Fig. \ref{crpt_fluctuations}).

For $p^*<p<1$, the failure sequence is non-trivial. In this case, some of the stronger fibers may break before some of the weaker fibers, given that 
the stronger fibers received load initially but did not break. They are now breaking, because some of the weaker fibers have broken and that increases the
load on them gradually. The basic point is that the fibers now are carrying loads close to their thresholds up to the point it breaks (load increment steps
are rather small). This increases the ``efficiency" of the system in a similar way seen in Ref. \cite{prl15}. The non-monotonicity of the phase boundary is, therefore,
a direct consequence of the non-uniformity of the initial loads. Again for $p\approx 1$ we checked that the failure sequence is in the order of the individual thresholds.

%%%%%%%%%%%%%%%%%%%%%%%%%%%%%%%%%%%%%%%%%%%%%%%%%%%%%%%%%%%%%%%%%%%%%%%%%%%%%%%%%%%%%%%%%%%%%%%%%%%%%%%%%%%%%%%%%%%%%%%%%%%%%%%%%%%%%%%%%
\begin{figure}[tbh]
\centering 
%\captionsetup{justification=raggedright}
\includegraphics[height=7cm]{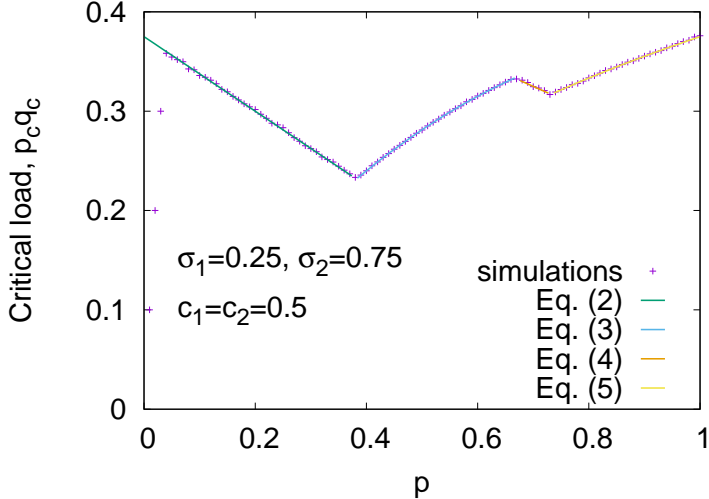}
   \caption{The phase boundaries for the bi-modal threshold distribution version of the mean field model are shown for a given set of parameters. 
The simulation points are matching the analytical results precisely. The only departures in the small $p$ limit is due to system size limitation.}
\label{phdia_all_lines}
\end{figure}
%%%%%%%%%%%%%%%%%%%%%%%%%%%%%%%%%%%%%%%%%%%%%%%%%%%%%%%%%%%%%%%%%%%%%%%%%%%%%%%%%%%%%%%%%%%%%%%%%%%%%%%%%%%%%%%%%%%%%%%%%%%%%%%%%%%%%%%%%%
To check if the non-monotonic variation of the critical load is independent of the threshold distribution, we have simulated the system for
triangular and Gaussian distributions as well, both centered at 0.5. Unlike the other distributions considered here, Gaussian distribution
is essentially defined for all positive values of its argument. As can be seen from Fig. \ref{gaussian} the failure threshold distribution shows the
similar qualitative behavior of being non-monotonic with the fraction of the system initially loaded.

\subsection{Bi-modal threshold distribution}

To understand the non-monotonic behavior of the failure threshold with fraction of initially loaded system quantitatively, we consider 
a version of the model, where there are just two groups of fibers with relative probabilities $c_1$ and $c_2$ and having failure thresholds
$\sigma_1$ and $\sigma_2$, with $\sigma_1<\sigma_2$. As before, $p$ fraction of the system is loaded with load $q$.  For a particular set of parameters $c_1=c_2=0.5$, $\sigma_1=0.25$ and $\sigma_2=0.75$, all the boundaries
were evaluated and were matched with simulations (see Fig. \ref{phdia_all_lines}). The correspondence is 
perfect. The particular scenarios are detailed below:

{\bf Scenario 1:} If $q$ is so large that all the fibers to which the initial load is applied, are broken, i.e. $q>\sigma_2$. 
Then the total load is redistributed in the first step to all the remaining fibers, with load per fiber value $pq/(1-p)$. For the
dynamics to continue, it must satisfy $pq/(1-p)>\sigma_1$, breaking all the weaker fibers. Finally, the total load falls upon 
the stronger fibers, which did not receive any load on the first step, giving a load per fiber value $pq/(c_2(1-p))$. The critical
condition is when this becomes equal to the higher threshold:
\begin{equation}
p_cq_c=\sigma_2c_2(1-p_c).
\label{line1}
\end{equation}

{\bf Scenario 2:} If $\sigma_1<q<\sigma_2$, then initially all the weak fibers receiving load will break. 
This will cause an increment of load per fiber value to all surviving fibers by an amount $q_1=pqc_1/((1-p)c_1+c_2)$,
which means $q_1$ load per fiber for the newly exposed fibers and $q+q_1$ load per fiber for the $pc_2$ fraction
of fibers (stronger). Now, if $q_1<\sigma_1$, but $q+q_1\ge \sigma_2$, then all the stronger fibers having higher
load break. Therefore, by now all the fibers that were exposed to the initial load have broken. The remaining
system will now have the uniform load per fiber value $pq/(1-p)$.  The system will break down completely when $pq/(1-p)>\sigma_1$,
causing the remaining weak fibers to break and then  $pq/c_2(1-p)\ge \sigma_2$, causing the remaining stronger fibers to break as well.
The second condition corresponds to the first critical line, and the inequality is therefore valid for higher values of $pq$, that we
consider in this scenario.  Then condition $pq/(1-p)>\sigma_1$, however, is a less stringent condition at this stage, and the 
critical line is given by the condition  $q+q_1\ge \sigma_2$,
which translates into
\begin{equation}
p_cq_c=\frac{p_c\sigma_2}{1+\frac{p_cc_1}{(1-p_c)c_1+c_2}}.
\label{line2}
\end{equation} 

{\bf Scenario 3:} As before, we start with $\sigma_1<q<\sigma_2$, but in the second step one can have $q_1\ge \sigma_1$, while
$q+q_1<\sigma_2$. Then all the weak fibers are eliminated from the system. The stronger fibers will bear loads $q_1+q_2$ or
$q+q_1+q_2$ depending upon the stage at which they were exposed to loading, with $q_2=\frac{p(1-p)qc_1^2}{(1-p)c_1c_2+c_2^2}$.
The system will collapse completely when $q+q_1+q_2\ge \sigma_2$ and $pq/c_2(1-p)>\sigma_2$. If we put an equality on the second
condition, it gives the first critical line, but $pq$ here is higher here, so the inequality is trivially satisfied. The most stringent 
condition here is $q_1\ge \sigma_1$, which translates into:
\begin{equation}
p_cq_c=\frac{(1-p_c)\sigma_1c_1+\sigma_1c_2}{c_1}.
\label{line3}
\end{equation}

{\bf Scenario 4:} This is same as the scenario 3, except for the fact that the most stringent condition now becomes $q+q_1+q_2\le \sigma_2$,
which translates into:
\begin{equation}
p_cq_c=\frac{p_c\sigma_2}{1+\frac{p_cc_1}{(1-p_c)c_1+c_2}+\frac{p_c(1-p_c)c_1^2}{(1-p_c)c_1c_2+c_2^2}}.
\label{line4}
\end{equation}

%%%%%%%%%%%%%%%%%%%%%%%%%%%%%%%%%%%%%%%%%%%%%%%%%%%%%%%%%%%%%%%%%%%%%%%%%%%%%%%%%%%%%%%%%%%%%%%%%%%%%%%%%%%%%%%%%%%%%%%%%%%%%%%%%%%%%%%%%
\begin{figure}[tbh]
\centering 
%\captionsetup{justification=raggedright}
\includegraphics[height=7cm]{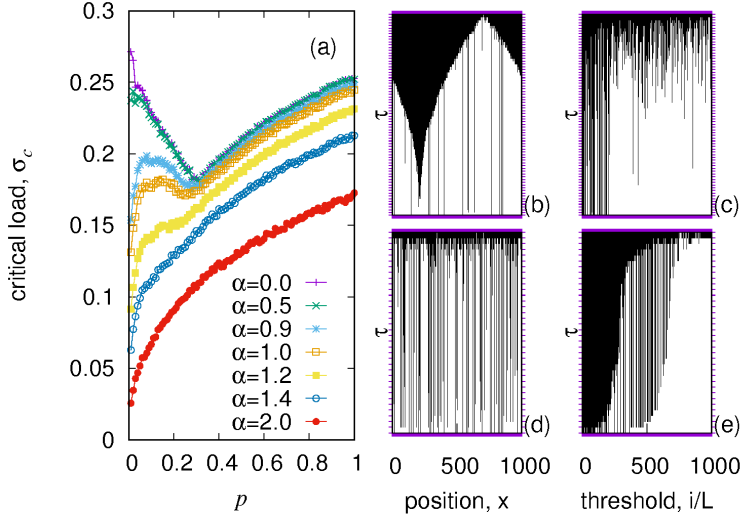}
   \caption{(a) The critical lines for different values of $\alpha$ for the one dimensional model. The non-monotonicity
disappears for more localized load distributions i.e. higher $\alpha$ values. Therefore, the correlations in the breaking 
sequence are either present in the real-space or in the threshold space. (b) and (d) shows the breaking sequence in the real
space, with $\alpha=2.0, p=0.6$ and $\alpha=0.1, p=0.6$ respectively as a function of the number of redistribution steps $\tau$
when the system is critically overloaded.
 For the more localized redistribution rule $\alpha=2$,
there is spatial correlation, but for $\alpha=0.1$ the breaking is random in real space. However, again for (c) and (e) 
 $\alpha=2.0, p=0.6$ and $\alpha=0.1, p=0.6$ respectively, but this time the breaking sequences are shown when the fibers
are arranged in ascending values of failure thresholds. Now for (e) a correlation is observed (as in the mean-field case),
that does not exist for (c).}
\label{sigmac_p_1d}
\end{figure}
%%%%%%%%%%%%%%%%%%%%%%%%%%%%%%%%%%%%%%%%%%%%%%%%%%%%%%%%%%%%%%%%%%%%%%%%%%%%%%%%%%%%%%%%%%%%%%%%%%%%%%%%%%%%%%%%%%%%%%%%%%%%%%%%%%%%%%%%%%
Finally, the crossover points between the lines can be obtained by assuming the continuity of the lines.
Specifically, continuity between Eq. \ref{line1} and Eq. \ref{line2} implies a crossover point
\begin{equation}
p^*=\frac{(1+c_2)-\sqrt{(1+c_2)^2-4c_1c_2}}{2c_1},
\label{cross1}
\end{equation}
the other solution is unphysical. Similarly, assuming continuity between 
Eq. \ref{line2} and Eq. \ref{line3}, the second crossover point can be obtained, which is
\begin{equation}
p^{**}=\frac{c_1(\sigma_1+\sigma_2)-\sqrt{c_1^2(\sigma_1+\sigma_2)^2-4c_1^2\sigma_1\sigma_2}}{2c_1^2\sigma_2},
\label{cross3}
\end{equation}
again the other solution is unphysical. Finally, continuity between Eq. \ref{line3} and Eq. \ref{line4},
will give the final crossover point.

%%%%%%%%%%%%%%%%%%%%%%%%%%%%%%%%%%%%%%%%%%%%%%%%%%%%%%%%%%%%%%%%%%%%%%%%%%%%%%%%%%%%%%%%%%%%%%%%%%%%%%%%%%%%%%%%%%%%%%%%%%%%%%%%%%%%%%%%%
\begin{figure}[tbh]
\centering 
%\captionsetup{justification=raggedright}
\includegraphics[height=7cm]{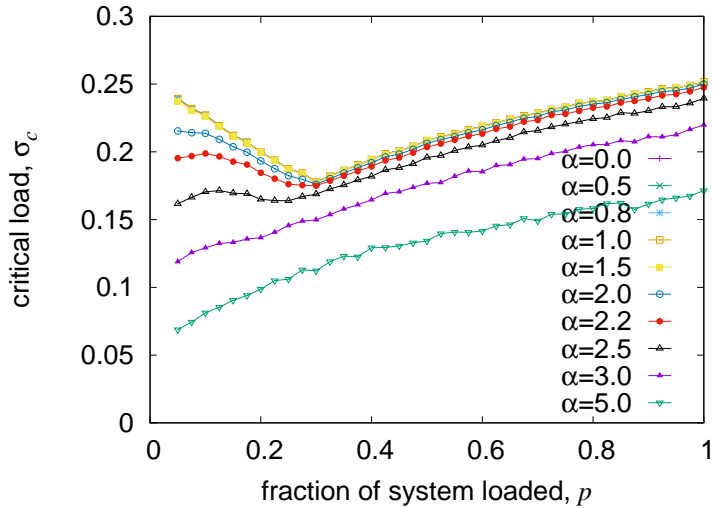}
   \caption{The variation of the critical load with initial loaded fraction for different values of $\alpha$ in the two dimensional model.}
\label{sigmac_p_2d}
\end{figure}
%%%%%%%%%%%%%%%%%%%%%%%%%%%%%%%%%%%%%%%%%%%%%%%%%%%%%%%%%%%%%%%%%%%%%%%%%%%%%%%%%%%%%%%%%%%%%%%%%%%%%%%%%%%%%%%%%%%%%%%%%%%%%%%%%%%%%%%%%%
\subsection{One dimensional model}
A more realistic case is the one dimensional model with power-law load sharing. In particular, a fiber at site $j$
will receive a load proportional to $\frac{1}{|i-j|^{\alpha}}$ when a fiber fails at $i$. Fig. \ref{sigmac_p_1d}(a) shows the variation of
the critical load (scaled for the partial loading condition) with the fraction of the system loaded for different values of $\alpha$.
The limits $\alpha\to 0$ and $p\to 0$ and 1 are the usual mean field uniform loading conditions. Other than that up to
$\alpha^*\approx 1$, there is a critical value for $p$ for which the critical load ($\sigma_c$) is minimum. For higher
values of $\alpha$, i.e. when the load sharing is very much local, the minimum critical load appears for the limit $p\to 0$.
In other words, for local load sharing limit of the model, the lowest critical load, for a given system size, is obtained when
the entire load is concentrated initially on one fiber. 

The reason for this can again be understood by looking at the breaking sequence of the fibers. Particularly, when the load sharing
is localized, the breaking of the fibers become spatially correlated. The local stress concentrations play a dominant role, rather than the 
order of the failure thresholds. We demonstrate that by looking at the failure sequence in the real space or when they
are arranged according to their failure thresholds. Note that the actual positions of the fibers are not changed at any time, 
just the failure sequence is noted according to their thresholds. In Figs. \ref{sigmac_p_1d}(d) and (e), we have $\alpha=0.1, p=0.6$.
In (d) the failure sequence is noted in the position space and there is no correlation. But in the threshold-ordered sequence 
(e), the correlation is clear and that is same as the one seen for the mean field case. Again, for (b) and (c), we have $\alpha=2.0, p=0.6$.
In this case, a spatial correlation is clear in (b), where as in threshold-ordered space (c) there is no correlation (except for
very early dynamics). 

\subsection{Two dimensional model}
The most general case is the above mentioned model executed for two dimensions. As before, a fiber at site $r_j$ will now receive
a load proportional to $\frac{1}{|r_i-r_j|^{\alpha}}$, when a fiber at $r_i$ breaks. Fig. \ref{sigmac_p_2d} shows the variation of critical load
as before. The minimum for a particular $p$ value now disappears around $\alpha^*\approx 2$ (see Fig. \ref{sigmac_p_2d}). For smaller values of
$\alpha$, the non-monotonic behavior of the failure threshold prevails. 

Therefore, even for the finite dimensional versions of the model, as long as the load redistribution is wide enough, there
exists a finite fraction for which is the critical load is minimum. This is possibly related to the fact that for broad enough 
load redistribution range, the finite dimensional model essentially behaves as the mean-field version. This is not entirely an
obvious limit for the model, since even when the mean of the range is non-divergent, the behavior can be mean field like. For
example, for two dimensions, such crossover happens for $\alpha^*\approx 2.25$, which is higher than the obvious limit $\alpha=2$ (see \cite{pre17,hidalgo02}).
Therefore, existence of the non-monotonicity is also expected for the similar limits of the model, which is important for fracturing of 
elastic solids, although the precise crossover values differ as the measures are different here.

It is also to be noted that for power law load sharing, the initially chosen fibers that are carrying the initial load are chosen
completely randomly. However, a spatial correlation in that choice, for example loading only a particular region of the system, can
have significant effect on the overall dynamics of the system. One specific limit was studied earlier in Ref. \cite{soc}, where the initial load
was applied only on one single centrally located fiber. Such spatial correlation and the possible correlation with the thresholds of the
initially chosen fibers, can be interesting questions for future investigations.

\section{Discussions and Conclusion}
The fiber bundle model with random failure thresholds is a rich prototypical model for failure of complex materials. 
Due to its simple and versatile nature, it has found application in modeling various systems, from composite solids
to power grids, road traffic etc. In this work we look at the failure strength of the model when the initial
load is applied to a finite fraction of the system. We find that for the mean field, one dimensional and two dimensional versions
of the model, the critical failure threshold or the load carrying capacity of the system is a non-monotonic
function of the initially loaded fraction, as long as the load redistribution function is sufficiently wide.
The non-monotonic variation remains qualitatively similar for different types of the threshold distributions of the fibers viz. uniform, triangular, Gaussian. 
The non-monotonic nature is suggestive in case of building redundancies for different systems. For example, 
it suggests that it is better to have the load distributed uniformly on all connections, if there are more than
one connections between two nodes carrying some load (e.g. power grids, computer networks etc.). On the other hand, 
to facilitate a breakdown, the same amount of load when applied to a critical fraction of the system causes more
damage than when applied uniformly or on a small fraction. It is also worth
noting at this point that the initial choice of the $p$ fraction of fibers was completely random in this case. 
However, a more educated guess, where at least some information about the failure thresholds of the individual
elements are taken into account (see e.g. \cite{prl15}), might improve the overall critical load. 

The common reason for the non-monotonic behavior is the non-monotonic failure sequence of the fibers with respect to their
respective thresholds. Particularly, a stronger fiber, if loaded initially, can break earlier than a weaker fiber that did not receive
any initial load. This reversal of breaking sequence is generally known to reduce the overall strength of the system (see, for e.g. Ref. \cite{prl15}
with $b>1$ case). At the onset of this non-monotonic breaking sequence, when the failure strength is minimum, the fluctuation of the
critical strength show a different scaling with system size as compared to the other values for the fraction of loading (see Fig. \ref{crpt_fluctuations}).
The basic mechanism of the failure strengths can be understood from a simple limit of the model (see Fig. \ref{phdia_all_lines}). However, this
mechanism breaks down when the load redistribution is very local. Then the failure probability due to stress concentration
becomes higher than any other fluctuations in the model. Nevertheless, the values of $\alpha^*$ are somewhat different from what
was found in Refs. \cite{pre17,biswas15} for which the behavior of the models shifts from local to global critical behavior. 
That is, however, a different measure and do not necessarily correspond with the non-monotonicity of the critical threshold studied here.

In conclusion, we find that for random fiber bundle models, the failure point of the system varies non-monotonically with 
the fraction of the system initially loaded, assuming a constant total load is applied to the system. This behavior is valid for
the mean-field, as well as the finite dimensional versions of the model. The results suggests that uniform application of the
load on the entire system is better than having a concentrated load. This can shed light on the design of redundancies in various
systems.

\section*{Acknowledgements}
PS thanks SERB project (Government of India)  for financial support. SB thanks
Humboldt foundation for financial support during part of the project.

\end{document}